\documentclass[%
 reprint,
 amsmath,amssymb,
 aps
]{revtex4-2}
\usepackage{graphicx}
\usepackage{dcolumn}
\usepackage{bm, xcolor}
\usepackage{subfigure, comment, float}

\usepackage[normalem]{ulem}

\begin{document}

\preprint{APS/123-QED}

\title{Phase Ordering in Binary Mixtures of Active Nematic Fluids}

\author{Saraswat Bhattacharyya}
 \email{saraswat.bhattacharyya@physics.ox.ac.uk}
\author{Julia M. Yeomans}%
 \
\affiliation{%
Rudolf Peierls Centre for Theoretical Physics, Parks Road, University of Oxford, OX1 3PU. United Kingdom.
}%


\begin{abstract}

We use a continuum, two-fluid approach to study a mixture of two active nematic fluids. Even in the absence of thermodynamically-driven ordering, for mixtures of different activities we observe turbulent microphase separation, where domains form and disintegrate chaotically in an active turbulent background. This is a weak effect if there is no elastic nematic alignment between the two fluid components, but is greatly enhanced in the presence of an elastic alignment or substrate friction.  We interpret the results in terms of relative flows between the two species which result from active anchoring at concentration gradients. Our results may have relevance in interpreting epithelial cell sorting and the dynamics of multi-species bacterial colonies.



\end{abstract}

\maketitle


\section{Introduction}

Phase separation is a ubiquitous phenomenon found across a wide variety of  biological systems. Inside cells, membrane-less organelles such as stress granules and nucleoli phase separate from their surroundings to allow different chemical environments for biochemical reactions \cite{Hyman2014, Banani2017}. Different cell types sort themselves into distinct regions in confluent layers \cite{KRENS2011, Suzuki2021} during growth and morphogenesis \cite{Sharrock2020, Hogan1999}. Bacterial colonies also undergo segregation, with species of different phenotypes clustering together \cite{Nadell2016}.    

A large body of research has looked at biological phase separation through the lens of equilibrium thermodynamics, attributing the ordering to the minimization of free energy. A notable thermodynamic model for cell sorting is the differential adhesion hypothesis (DAH) \cite{Steinberg1975, Foty2005},  which proposes that cells preferentially adhere to other cells of the same type because of differences in surface tension between like and unlike cells.  Other thermodynamic approaches include considerations of line tension \cite{Brodland2002, KRENS2011} and surface contraction \cite{Harris1976, KRENS2011} of cells.
However, biological matter is inherently out of thermodynamic equilibrium - which opens the possibility of phase ordering mechanisms that are outside the realm of free energy minimization principles \cite{Heine2021}. 

An important class of non-equilibrium systems is active matter, which deals with the collective behaviour of self-motile particles. Motility-induced phase separation (MIPS) \cite{Cates2015, Gonnella2015} is an example of active phase separation, where self-propelled particles can become trapped in regions of high density thus forming a dense phase and a dilute phase.  Many other novel mechanisms of ordering in active systems have been reported in the literature, including aligning torques \cite{Zhang2021}, bond formation and breaking \cite{Gokhale2022}, control of boundary conditions \cite{Thutupalli2018}, and the hydrodynamic interactions between dumbbell-shaped swimmers \cite{Furukawa2014}. Continuum models of scalar active matter have been used to study phase separation in active Brownian particles \cite{Stenhammer2013}, self-propelled particles \cite{Wittkowski2014, Tiribocchi2015}, poroelastic materials \cite{Weber2018} and cellular aggregates \cite{Kuan2021}. These show steady states that range from bulk phase separation to bubbles, droplets, elongated filaments and active foams. Recent work on active phase field models \cite{Balasubramaniam2022, Graham2024} and a vertex model \cite{Krajnc2020} on mixed cell layers have also shown phase separation.

Active nematics \cite{SimhaRamaswamy2002, MarchettiReview, Doostmohammadi2018} comprise rod-like particles with orientational order, which pump energy into their surroundings by generating dipolar stresses along their long axes. These models have been successfully used to describe the motility of Madin-Darby Canine Kidney (MDCK) cells \cite{Saw2017}, spontaneous flow in confined cell channels \cite{Duclos2018}, active turbulence in microtubule-kinesin mixtures \cite{Opathalage2019}, and topological defects in growing bacterial colonies \cite{DellArciprete2018}. Recently, Assante et. al. \cite{Assante2023} showed that coupling concentration and nematic ordering can lead to spontaneous microphase separation in inhomogeneous active nematics, and we used a continuum theory to study active phase separation, driven by flows, in a mixture of an active nematic and a passive isotropic fluid \cite{Bhattacharyya2023}. 
In this paper, we extend this  work to discuss mixtures of two nematics, with different activities, coupled by viscous drag. This is motivated by recent experiments which demonstrate cell sorting in mixtures of extensile and contractile cells \cite{Balasubramaniam2022, Skamrahl2023}. 

The paper is organized as follows: In Sec.~2, we extend the active two-fluid model introduced in Ref. \cite{Bhattacharyya2023} to describe interacting active nematics. In Sec.~3, we discuss the angle between the orientations of the two active species when they are coupled only by viscous drag. In Sec.~4, we move on to phase separation in active-active mixtures. We review the mechanism discussed in Ref. \cite{Bhattacharyya2023}, and show how this applies when both species are active, and how the phase ordering depends on the elastic coupling between nematogens. In Sec.~5, we discuss how changing concentration fractions and friction affect phase separation in an extensile-contractile mixture. 
Finally, in Sec.~6, we conclude with a summary of our results.

\section{Model}

We study a mixture of two active fluids \cite{levy1999, Joanny2007, Weber2018, Bhattacharyya2023}.  Each fluid component has a local density $\rho^i$, velocity field $u_\alpha^i$, and chemical potential $\mu^i$, where Latin superscripts $i=1, 2$ index the different fluids, and the Greek subscripts $\alpha = 1, 2 $ denote the spatial directions. Summation convention is used for the Greek indices but will be specified explicitly for the Latin indices when applicable. 

Each component fluid obeys the mass continuity equation
\begin{equation}
    \partial_t \rho^i + \nabla_\alpha \rho^i u_\alpha^i = 0
    \label{eqn:CompMassContinuity}
\end{equation}
and the momentum balance equation
\begin{equation}
    \begin{split}
        \partial_t \rho^i u_\alpha^i + \nabla_\beta \rho^i u_\alpha^i u_\beta^i = -\rho^i \nabla_\alpha \mu^i + F_\alpha^{visc, i} + F_\alpha^{body, i}\\ - f \phi^i u_\alpha^i+\gamma \phi (1-\phi) (u_\alpha^{3-i} - u_\alpha^i)  ,
    \end{split}    
    \label{eqn:CompMomentumBalance}
\end{equation}
where
\begin{equation}
    \begin{split}
     \rho^c &= \rho^1 + \rho^2 \, ,\\
        \phi^1 &= \phi = \rho^1/\rho^c \, ,\\ 
        \phi^2 &= 1- \phi = \rho^2/\rho^c \,  
    \end{split}    
    \label{eqn:PhiVariables}
\end{equation}
are the total density, concentration fraction of component 1, and concentration fraction of component 2 respectively. In Eq.~(\ref{eqn:CompMomentumBalance}), the left-hand side denotes the convective derivative of the fluid momentum density ($\rho^i u_\alpha^i$), while the right-hand side describes the force acting on the fluid per unit volume. The forces are modelled by a thermodynamic force ($-\rho^i \nabla_\alpha \mu^i$), a viscous drag between the component fluids $\gamma \phi (1-\phi) (u_\alpha^{3-i} - u_\alpha^i)$, an internal viscous dissipation for each fluid $F_\alpha^{visc, i}$, a substrate friction term $-f\phi^iu_\alpha^i$, and a body force $F_\alpha^{body, i}$ which models the local forces generated by nematic stresses.

We formulate our equations to treat this system as an incompressible fluid with compressible components. In order to do so, we define new velocity fields 
\begin{equation}
    \begin{split}        
        u_\alpha^c &= \phi u_\alpha^1 + (1-\phi) u_\alpha^2 \, ,  \\ 
        \delta u_\alpha &= u^1_\alpha - u^2_\alpha ,
    \end{split}    
    \label{eqn:CombinedVariables}
\end{equation}
which are the centre of mass velocity of the total fluid and relative flow between the fluids respectively. We reserve the superscript $c$ to refer to the combined (centre of mass) fluid. Moving forward, we will assume that the relative flow is much smaller than the combined velocity of the fluid i.e. $|\delta \mathbf u| \ll |\mathbf u^c|$ .

Adding Eqs.~(\ref{eqn:CompMassContinuity}), (\ref{eqn:CompMomentumBalance}) for each component, and neglecting terms of order $(\delta \mathbf{u})^2$, gives the equations of motion for the combined fluid \cite{Malevanets1999, Malevanets2000}:

\begin{equation}
    \partial_t \rho^c + \nabla_\alpha \rho^c u_\alpha^c = 0,
    \label{eqn:CombiMassContinuity}
\end{equation}
\begin{equation}
    \begin{split}
    \partial_t \rho^c u_\alpha^c + \nabla_\beta \rho^c u_\alpha^c u_\beta^c &= \sum_{i=1}^{2} \big[ -\rho^i \nabla_\alpha \mu^i  +  \\ & F_\alpha^{visc, i}  -f\phi^i u_\alpha^i + F_\alpha^{body, i}  \big].
    \label{eqn:CombiMomentumBalance}
    \end{split}
\end{equation}
Notice that the viscous drag between the components drops out of the equation for the combined fluid. The combined fluid conserves mass density and momentum and acts as a typical incompressible fluid. 

We now discuss the terms on the right-hand side of the momentum balance equation~(\ref{eqn:CombiMomentumBalance}) in turn. The first term is the thermodynamic force, which follows from a Ginzburg-Landau free energy functional of the form \cite{Malevanets1999, Malevanets2000} 
\begin{equation}
    \mathcal{F}_{LG} = \int d^2\mathbf{r} \, \big(\psi(\rho^1, \rho^2) + \frac{1}{2} \kappa || \mathbf{\nabla}\phi ||^2 \big)
    \label{eqn:LandauGinzburgFE}
\end{equation}
where 
\begin{equation}
    \psi = \frac{1}{3}\rho^c  \ln \rho^c  + \rho^c \{ a \,(\phi - \frac{1}{2})^2 + b\,(\phi - \frac{1}{2})^4 \}  .
    \label{free}
\end{equation}
The first term of~(\ref{free}) promotes  incompressibility of the combined fluid, with an isothermal equation of state. The second term is a Landau free energy which drives the system to a uniformly mixed configuration if $a \geq 0, b \geq 0$. The chemical potentials for each fluid are defined as $ \mu^i = \partial \mathcal{F}_{LG}/\partial \rho^i $. We can write \cite{Malevanets1999, Malevanets2000} the chemical potential term for the combined fluid $\sum_{i=1}^{2} -\rho^i \nabla_\alpha \mu^i$ as the divergence of a stress tensor $-\nabla_\beta\sigma^{thermo}_{\alpha\beta}$, where
\begin{equation}
    \sigma^{thermo}_{\alpha\beta} = p \delta_{\alpha\beta} + \kappa(\nabla_\alpha \nabla_\beta \phi - \frac{1}{2} || \mathbf{\nabla} \phi ||^2 \delta_{\alpha\beta} ) ,
    \label{eqn:LG_stress}
\end{equation}
\begin{equation}
    -p = \psi - \frac{\partial \psi}{\partial \rho^1} \rho^1 - \frac{\partial \psi}{\partial \rho^2} \rho^2 .
\end{equation}
$p = \rho^c/3$ is an isotropic pressure consistent with an isothermal equation of state. The term in $\kappa$ is an anisotropic stress resulting from the surface tension between the two component fluids. 

The second term on the right-hand side of Eq.~(\ref{eqn:CombiMomentumBalance}) is the usual viscous stress, defined by 
\begin{equation}
    F_\alpha^{visc, i} = \nabla_\beta \, \eta^i (\nabla_\alpha u^i_\beta + \nabla_\beta u^i_\alpha - \delta_{\alpha\beta}\nabla_\gamma u^i_\gamma ).
    \label{eqn:ViscousStress}
\end{equation}
The third term describes the friction between each component fluid and the substrate.
The final term in Eq.~(\ref{eqn:CombiMomentumBalance}) is the body force acting locally on the fluid at each point which arises from passive and active nematic stresses  $  F_\alpha^{body, i} \equiv F_\alpha^{Q, i} $. We next discuss the dynamics of each active nematic species and the form of $F_\alpha^{Q, i} $.

Each fluid component $i$ is a nematic liquid crystal \cite{SimhaRamaswamy2002, MarchettiReview, Doostmohammadi2018} described by  a symmetric traceless tensor \cite{degennes_book} 
\begin{equation}
    Q^i_{\alpha\beta} =  S^i_{nem} (2 n^i_\alpha n^i_\beta - \delta_{\alpha\beta})
    \label{eqn:Qi}
\end{equation}
in 2D where $n^i_\alpha$ is a headless vector denoting the orientation of the local nematic order, called the director field, and $S^i_{nem}$ is the magnitude of the nematic order. The order parameter relaxes towards the minimum of the Landau-de Gennes free energy density \cite{Doostmohammadi2018, degennes_book} 
\begin{equation}
    \mathcal{F}^{LdG, i} = C^i (\, (S^i_{nem})^2 - Q^i_{\alpha\beta}Q^i_{\alpha\beta})^2 + \frac{K^i}{2} \nabla_\gamma Q^i_{\alpha\beta} \nabla_\gamma Q^i_{\alpha\beta}
    \label{eqn:LdG}
\end{equation}
defined so that the minimum free energy corresponds to a state with an order parameter of magnitude $S_{nem}^i$, with the director uniformly aligned in space.

We couple the orientation fields of the two species explicitly by adding an extra aligning or anti-aligning interaction in the free energy of the form 
\begin{equation}
    \mathcal{F}^{Q_1,Q_2} = -L_{12} \phi (1-\phi) \,Q_{\alpha\beta}^{1}\, Q_{\alpha\beta}^{2} .
    \label{eqn:CrossElasticity}
\end{equation}
When both species are mixed together, this term aligns the director fields for $L_{12}>0$, and anti-aligns them for $L_{12}<0$. From Eqs.~(\ref{eqn:LandauGinzburgFE}), (\ref{eqn:LdG}) and (\ref{eqn:CrossElasticity})
the total free energy of the system is given by
\begin{equation}
    \mathcal{F} = \mathcal{F}_{LG} + \mathcal{F}^{LdG, 1} + \mathcal{F}^{LdG, 2} + \mathcal{F}^{Q_1,Q_2}.
    \label{eqn:totalFreeEnergy}
\end{equation}

We now describe the dynamics of the orientation field. We expect the director field to not only be advected by the fluid but also to be rotated according to the vorticity tensor $\Omega^i_{\alpha\beta} = (\partial_\alpha u^i_\beta - \partial_\beta u^i_\alpha)/2$. Additionally, the director may tend to orient along the strain rate tensor $E^i_{\alpha\beta} = (\partial_\alpha u^i_\beta + \partial_\beta u^i_\alpha)/2$ of the fluid. The response to gradients of the flow is modelled by the co-rotation term \cite{degennes_book, Doostmohammadi2018}
\begin{equation}
    \begin{aligned}
    S_{\alpha\beta}^i &= (\lambda E^i_{\alpha\chi} + \Omega^i_{\alpha\chi}) (Q^i_{\chi\beta}+\delta_{\chi\beta}/2 ) + \\
    & \qquad \qquad(Q^i_{\alpha\chi}+\delta_{\alpha\chi}/2 )(\lambda E^i_{\chi\beta} - \Omega^i_{\chi\beta}) \\
    & \qquad \qquad-  2\lambda^i (Q^i_{\alpha\beta}+\delta_{\alpha\beta}/2) Q^i_{\chi\gamma}\nabla_{\chi}u^i_\gamma .
    \end{aligned}
    \label{eqn:Corot_term}
\end{equation}
where $\lambda^i$ (known as the flow-tumbling or flow-aligning parameter) models the extent to which the director field of species $i$ explicitly orients with the strain axis of the fluid (the direction of the positive eigenvalue of $E^i_{\alpha\beta}$).

We also expect that the system will evolve towards the minimum of the free energy. This is modelled by the molecular field  \cite{Doostmohammadi2018, SimhaRamaswamy2002} 
\begin{equation}
    H_{\alpha\beta}^i =-\frac{\partial \mathcal{F}}{\partial Q^i_{\alpha\beta}} + \frac{\delta_{\alpha\beta}}{2} \bigg( \frac{\partial \mathcal{F}}{\partial Q^i_{\chi\gamma} } \bigg) \delta_{\chi\gamma} . 
    \label{eqn:mol_Field}
\end{equation}
Combining Eqs.~(\ref{eqn:Corot_term}) and (\ref{eqn:mol_Field}), the dynamics of the nematic tensor associated with fluid $i$ is \cite{Beris1994}
\begin{equation}
    \partial_t Q^i_{\alpha\beta} + u^i_\chi \nabla_\chi Q^i_{\alpha\beta} - S^i_{\alpha\beta} = \Gamma H^i_{\alpha\beta}.
    \label{eqn:Q_Field}
\end{equation}

The nematic field itself generates stresses, which drive flows in the fluid. This is modelled by a body force 
\begin{equation}
    F^{Q, i}_\alpha = \nabla_\beta \cdot \Pi^i_{\alpha\beta}
    \label{eqn:ForceQ}
\end{equation}
where the stress tensor is a sum of the elastic and active stresses \cite{Doostmohammadi2018, MarchettiReview} 
\begin{flalign}
  \Pi^i_{\alpha\beta}       &= \Pi_{\alpha\beta}^{i, el} + \Pi_{\alpha\beta}^{i, act} \label{eqn:TotalStress},\\
    \Pi^{i, el}_{\alpha\beta}   &= 2\lambda (Q^i_{\alpha\beta} + \delta_{\alpha\beta}/2)\,Q^i_{\chi\gamma}H^i_{\chi\gamma} -  \lambda H^i_{\alpha\chi} (Q^i_{\chi\beta} + \delta_{\chi\beta}/2) \nonumber  \\  
    & \qquad- \lambda (Q^i_{\alpha\chi} + \delta_{\alpha\chi}/2)H^i_{\chi\beta} - \nabla_\alpha Q^i_{\chi\gamma} \frac{\partial \mathcal F^{LdG}}{\partial \nabla_\beta Q^i_{\chi\gamma}}  \nonumber \\ &\qquad + Q^i_{\alpha\chi} H^i_{\chi\beta} - H^i_{\alpha\chi} Q^i_{\chi\beta}, \label{eqn:ElasticStress} \\
    \Pi^{i, act}_{\alpha\beta} &= -\zeta_i \phi^i Q^i_{\alpha\beta} \label{eqn:ActiveStress}.  
\end{flalign}

Now, we return to the equation of motion for the first compressible fluid component.  
Using Eqs.~(\ref{eqn:CompMassContinuity}) and~(\ref{eqn:CombinedVariables}), the momentum equation~(\ref{eqn:CompMomentumBalance}) for the first  component can be written as
\begin{equation}
    \begin{split}
    \rho^1 [ \partial_t  u_\alpha^1 + u_\beta^1 \nabla_\beta  u_\alpha^1 ]  = -\phi \nabla_\beta \sigma_{\alpha\beta} + F_\alpha^{visc, 1} + F_\alpha^{body, 1} \\ - f\phi u_\alpha^1+ [\gamma \phi (u_\alpha^c - u_\alpha^1)+\phi (1-\phi)\nabla_\alpha \delta \mu]  
    \end{split}
    \label{eqn:CompFinalMomentumBalance}
\end{equation}
where $\delta \mu = -\delta\mathcal{F} /\delta \phi$ \cite{Malevanets1999, Malevanets2000}.  Defining an internal force density acting between the fluids,
\begin{equation}
    G_\alpha = \gamma (u_\alpha^c - u_\alpha^1)+ (1-\phi)\nabla_\alpha \delta \mu,
    \label{eqn:G_force}
\end{equation}
Eq.~(\ref{eqn:CompFinalMomentumBalance}) can then be rewritten as
\begin{equation}
        \rho [ \partial_t  u_\alpha^1 + u_\beta^1 \nabla_\beta  u_\alpha^1 ]  = - \nabla_\beta \sigma_{\alpha\beta} + G_{\alpha} + \frac{F_\alpha^{visc, 1} + F_\alpha^{body, 1} }{\phi} - fu_\alpha^1.
    \label{eqn:CompFinalMomentumBalance_WeirdVersion}
\end{equation}
The equation of the second compressible fluid component follows by symmetry.

We numerically solve the combined fluid, described by Eqs.~\eqref{eqn:CombiMassContinuity}-\eqref{eqn:CombiMomentumBalance}, using the Lattice-Boltzmann (LB) method \cite{LB_Book}. We simultaneously calculate the evolution of the first component fluid, described by Eqs.~(\ref{eqn:CompMassContinuity}) and (\ref{eqn:CompFinalMomentumBalance}), using a LB method modified to account for compressibility \cite{Malevanets1999, Malevanets2000}. The dynamics of the nematic fields and stresses (Eqs.~(\ref{eqn:Q_Field})-(\ref{eqn:ActiveStress})) are solved using a finite difference approach \cite{Blow2014}. 

We used the parameters $\rho^C = 40$, $\gamma = 4$, $\kappa = 5$, $a = 0.0001$, $b= 0.0001$, $\eta^1 = \eta^2 = 10/3$, $S_{nem}^1 = S_{nem}^2 = 1$, $C^1 = C^2 = K^1=K^2 = 0.1$,  $L_{12} = 0.1$, $\lambda=0$, $f=0$, and $ \Gamma^1 = \Gamma^2 =0.1$, unless otherwise specified. $\zeta_1, \zeta_2$ were varied in the range $[-0.1, 0.1]$.
We simulated  the equations for 50,000 time steps on a 200 x 200 grid with periodic boundary conditions. The initial condition was chosen to be $\rho^1 = \rho^2 = 20$, with the initial director and velocity field configurations obtained by simulating 500 LB time-steps from a randomly initialized director configuration without active forcing.  

\section{Active flows introduce director alignment between nematic fluid components}
\label{sec:FlowDrivenAlignment}

\begin{figure*} [thp]
    \centering    
    \includegraphics[width=\textwidth]{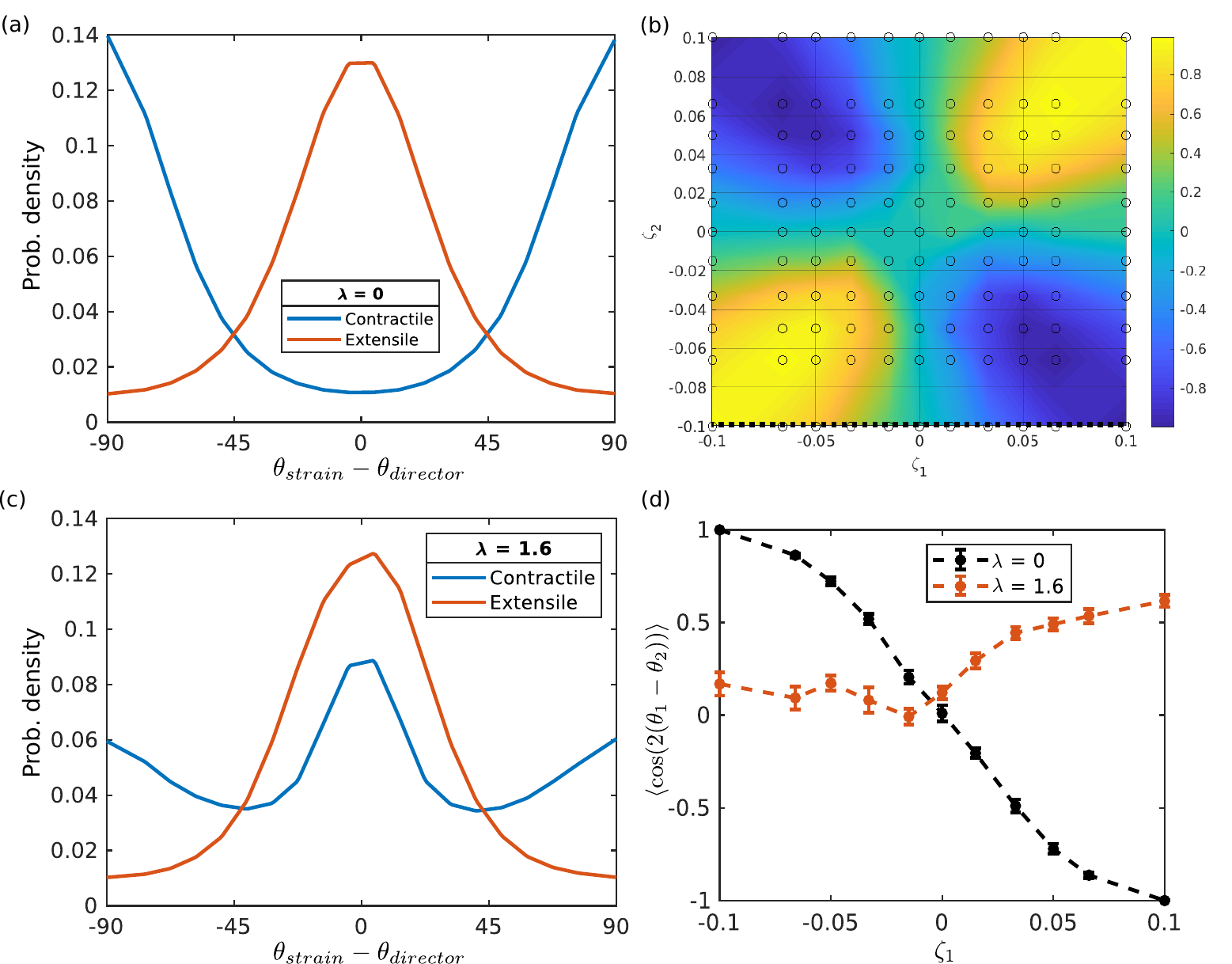}
 \caption{\textbf{Alignment between nematic director fields.} (a) Probability density function shows that for $\lambda=0$, contractile nematics tend to align perpendicular to the extensional strain axis (blue), while extensile nematics align parallel (red) ($\zeta_1 = -0.10, \zeta_2 = 0.067$). (b) Contour plot showing nematic alignment between the different components for different activities for $\lambda=0$. The colourbar shows $\langle\cos(2(\theta_1-\theta_2))\rangle$ which is +1 (yellow) for parallel and -1 (blue) for perpendicular. Circles denote individual simulations, while the contour plot shows the interpolated values. (c) Probability density function shows that for $\lambda=1.6$, contractile nematics have a bimodal distribution preferentially aligning either parallel and perpendicular to the extensional strain axis (blue), while extensile nematics tend to align parallel (red) ($\zeta_1 = -0.10, \zeta_2 = 0.05$). (d) Cross-section of nematic alignment while varying $\zeta_1$, with $\zeta_2 = -0.10$. The $\lambda=0$ cross-section is marked by the black dotted line in the contour plot. Changing to $\lambda=1.6$ changes the nematic alignment significantly in both contractile-contractile ($\zeta_1<0$) and contractile-extensile ($\zeta_1>0$) mixtures. }
    \label{fig:Alignment}
\end{figure*}

In this section, we argue that there is alignment (or anti-alignment) between the two components of the nematic mixture due to the director fields aligning with the rate of strain tensor in the combined fluid. Throughout this section, we set $L_{12} = 0$ and $a= 0.2$ to study flow-induced director-director coupling in the absence of imposed elastic alignment or  phase separation. We first consider the flow tumbling parameter $\lambda=0$, and then discuss the effects of a non-zero $\lambda$. 

\subsection{Zero flow-tumbling parameter ($\lambda=0$)}

Assuming that the momentum balance in each fluid  (Eq.~(\ref{eqn:CompMomentumBalance})) is dominated by viscous and active stresses then, for $\lambda=0$,
 Eqs.~(\ref{eqn:ViscousStress}) and (\ref{eqn:ActiveStress}) give a force balance 
\begin{equation}
        \begin{split}
        \nabla_\beta\cdot \eta^i E^c_{\alpha\beta} = \zeta_{i} \nabla_\beta\cdot Q^{i}_{\alpha\beta} \qquad\\
    \end{split}    
    \label{eqn:SimplifiedStokes}
\end{equation}
for each of the fluid components. Here, we have assumed that both component fluids have approximately the same strain rate $E_{\alpha\beta}^c$.

If $\zeta_i>0$ (extensile), $Q_{\alpha\beta}$ has the same sign as $E_{\alpha\beta}$, and the nematic aligns along the stretching direction of the combined fluid ($\theta_{director} \parallel \theta_{strain}$). If $\zeta_i<0$ (contractile), $Q_{\alpha\beta}$ and $E_{\alpha\beta}$ have opposite signs, and the nematic aligns along the compression direction of the combined fluid ($\theta_{director} \perp \theta_{strain}$). This is shown in Fig.~\ref{fig:Alignment}(a). The spread in the distribution is due to elasticity in the nematic field of each component and the effect of the passive backflow terms in the Navier Stokes equations which are neglected in Eq.~(\ref{eqn:SimplifiedStokes}). \\

When two active fluids are strongly coupled through viscous drag, the alignment of each nematic with the combined fluid velocity gradient induces an effective alignment between the two director fields. Both species orient in the same (perpendicular) direction if they have the same (opposite) sign of activity. The overall degree of alignment can be characterised by calculating the average value of $\cos(2(\theta_1-\theta_2))$, where $\theta_i$ is the angle of the $i$-th director field. This quantity is +1 for parallel alignment and -1 for perpendicular alignment as shown in Fig.~\ref{fig:Alignment}(b).

\subsection{Effect of the flow-tumbling parameter $\lambda$}

We now consider the effects of having a large flow-tumbling parameter $\lambda$. The passive elastic terms proportional to $\lambda$ in the stress (Eq.\eqref{eqn:ElasticStress}) tend to align the nematic director field with the extensional strain axis, regardless of active flows, for both extensile and contractile nematics. For an extensile fluid this merely slightly enhances the alignment along the strain axis caused by the activity. However, the angle between contractile nematogens and the strain axis forms a bimodal distribution with some regions aligning parallel to the strain, due to elastic flows, and others perpendicular to the strain, due to active flows, with the relative fraction of each alignment dependent on $\lambda$.
(Fig. \ref{fig:Alignment}(c)).

This in turn can affect the effective alignment between the director fields of the two components. For extensile-extensile mixtures, there is just a small decrease in $\langle  \theta_1 - \theta_2 \rangle$.
For an extensile-contractile mixture,  some contractile directors align with the extensional strain axis, leading to an increase in alignment between the two species. For a contractile-contractile mixture, both director fields are frustrated, and $\langle  \theta_1 - \theta_2 \rangle$ depends on the fraction of director fields which align parallel or perpendicular to the strain. In Fig. \ref{fig:Alignment}(d), we plot the average nematic alignment $\langle\cos(2(\theta_1-\theta_2))\rangle$ along a cross-section varying $\zeta_1$ keeping $\zeta_2 = -0.10$ fixed, for both $\lambda=0$ and $\lambda=1.6$.

\section{Phase Separation in Active-Active Mixtures}

In previous work, we showed that an active nematic fluid mixed with a passive isotropic fluid spontaneously orders to form microphase-separated domains \cite{Bhattacharyya2023}, even in the absence of any terms in the free energy favouring phase separation. The domains form and disintegrate chaotically in an active turbulent background, a state we  term `turbulent microphase separation'. 

In this section, we first review the phase-separation mechanism and then apply it to the case of two active nematic fluids of different activities. Secondly, we determine  how the strength of phase separation depends on the activities of each species.  Finally, we look at how the results change when the nematic species are constrained to align with each other through elastic interactions. 

\subsection{Mechanism}

\begin{figure*}[htp]
    \centering    
    \includegraphics[width=\textwidth]{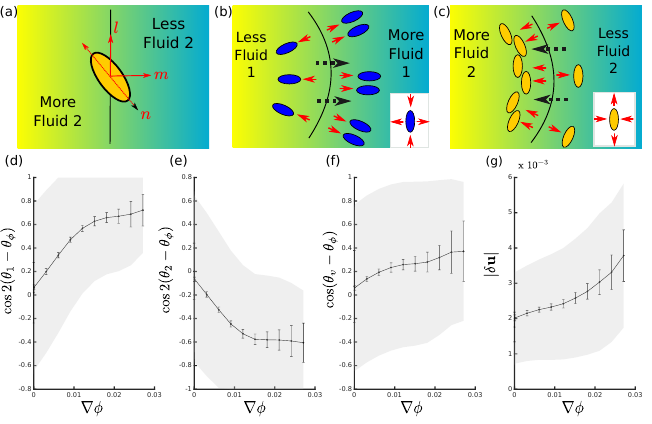}
    \caption{\textbf{Mechanism of Phase Separation:} (a) Interface normal $\mathbf{m}$, tangent $\mathbf{l}$, and nematic director $\mathbf{n}$
     at an interface. Schematic representation of the mechanism of phase separation for (b) a contractile component 1 (blue)  and (c) an extensile component 2 (yellow). Red arrows show the forces at the interface created by each nematogen, and black arrows show the net direction of the active force. (d)--(f) simulation data demonstrating the mechanism for $\zeta_1 = -0.10, \zeta_2 = 0.067$, and $ L_{12}=0$: (d) Contractile species align homeotropically at interfaces. (e) Extensile species align parallel to the interface. (f) The relative flow between the fluid components tends to orient normal to the interface, in the direction of increasing concentration gradient. (g) Relative flows between the fluids are stronger at the interface. The trend lines show the mean, combining data from 50 measurements taken in intervals of 1000 timesteps. The error bars indicate the standard deviation of the mean. The shaded region shows the standard deviation in the spread of the measured quantities.}    \label{fig:Mechanism}
\end{figure*}

Active phase separation begins when the two component species generate a flow fluctuation which locally moves them in different directions, setting up a small concentration gradient in the fluid. This difference in concentration leads to an imbalance in active stress across the interface, which drives active flows normal and tangential to the interface in each fluid. 

Consider how each active fluid component $i$ behaves at an interface. This fluid generates tangential and normal active forces per unit volume given by $ \mathbf{F}^{\footnotesize{\mbox{tangential}, i}} = 2 \zeta_i  |\nabla (S_{nem}^i\phi^i)| \,(\mathbf{m \cdot n}) (\mathbf{l \cdot n}) \,\mathbf{l} $ and $ \mathbf{F}^{\footnotesize{\mbox{normal}, i}} = -\zeta_i  |\nabla (S_{nem}^i \phi^i)| \, (2 (\mathbf{m \cdot n})^2 -1 ) \,\mathbf{m}$ respectively \cite{Bhattacharyya2023, Blow2014}. Here $\mathbf{m}$ and $\mathbf{l}$ are unit vectors normal (pointing away from the more active region) and  tangential to the interface respectively, and $\mathbf{n}$ is a unit vector along the nematic director as shown in Fig. \ref{fig:Mechanism}(a).

The tangential flows acting on the fluid component tend to orient its director parallel to the interface for extensile nematics, and perpendicular to the interface for contractile nematics \cite{Blow2014}. Due to this {\em active anchoring}
the active stresses generated by each fluid component, and acting on that component, tend to point normal to the interface and towards the region of higher concentration, as shown by the black arrows in Fig.~\ref{fig:Mechanism}(b) for a contractile nematic, and Fig.~\ref{fig:Mechanism}(c) for an extensile one. 
Since both species are active, each fluid generates its own active stresses and flows. If the sum of the active forces is stronger than the passive restoring forces, this creates relative flows between the two components, magnifying the concentration difference across the interface further, leading to phase separation.
\begin{figure*}[thp]
    \centering    
    \includegraphics[width=\textwidth]{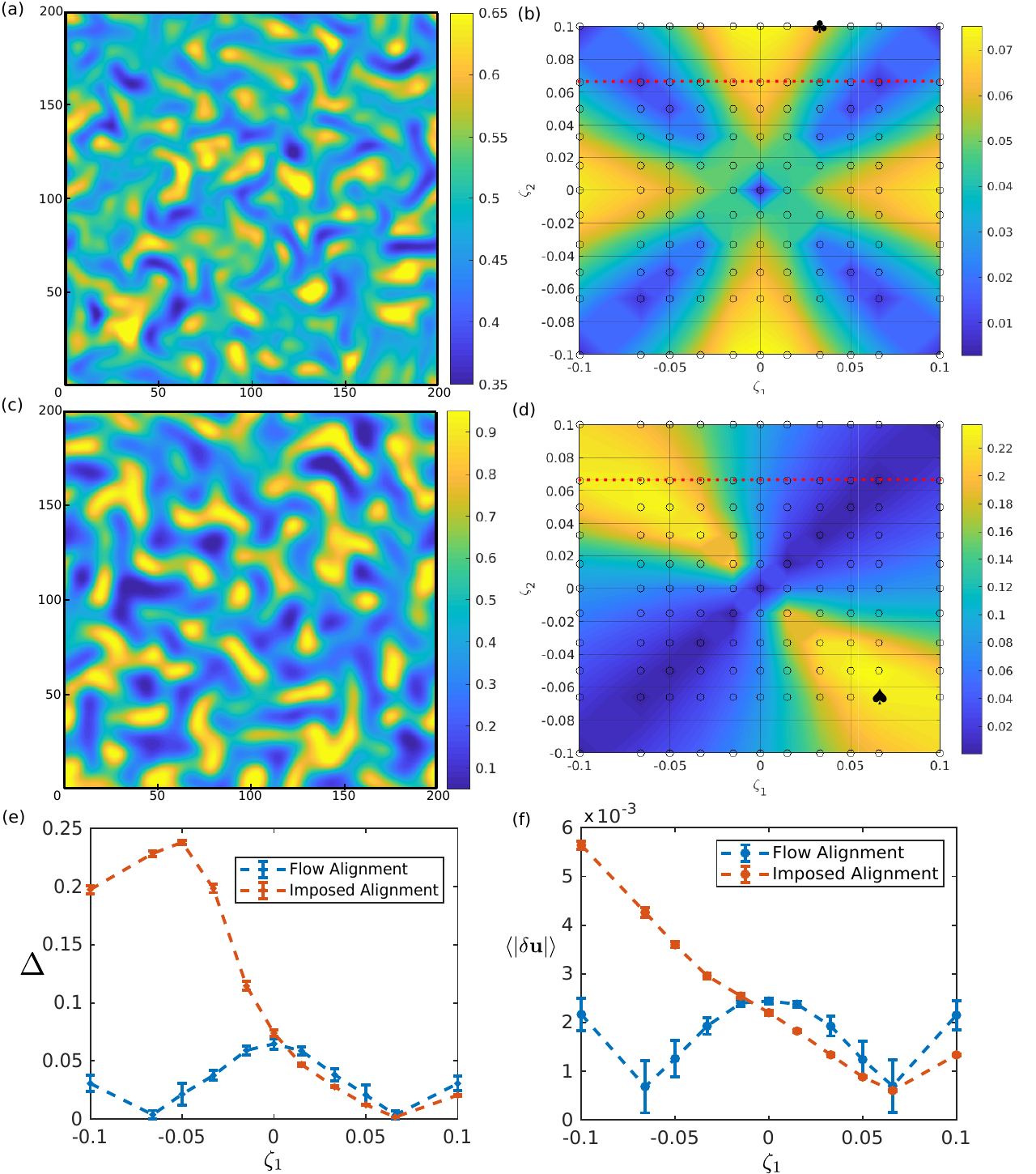}
    \caption{\textbf{Phase separation in active-active mixtures:}  (a) Snapshot of the concentration field for a turbulent microphase separated state without imposed alignment, $L_{12}=0$ (corresponding to the point marked by $\clubsuit$ in panel (b)).  The colourbar shows local concentration $\phi$. (b) Magnitude of the phase separation, $\Delta$,  for different activities. Circles denote individual simulations, while the contour plot shows the interpolated values. (c) Snapshot of the concentration field for a turbulent microphase separated state with imposed alignment, $L_{12}=0.1$ (corresponding to the point marked by $\spadesuit$ in panel (d)).  (d) Magnitude of the phase separation, $\Delta$, for different activities. Circles denote individual simulations, while the contour plot shows the interpolated values. Note the increase in the magnitude of the ordering when there is elastic alignment between the director fields of the fluid components. Variation of (e) the magnitude of the ordering, $\Delta$, and (f) the relative flow magnitude, $|\delta \mathbf u|$, with $\zeta_1$ for $\zeta_2=0.067$  (i.e.~along the red dotted lines in (b) and (d)), comparing $L_{12}=0$ and $L_{12} \neq 0$.
 }
    \label{fig:BasicPhaseSeparation}
\end{figure*}

We numerically verify this mechanism by looking at the active anchoring and flow alignment at concentration gradients.  We define the director angle of species $i$ as $\theta_i$, the orientation of the relative flow velocity $\delta\mathbf u$ as  $\theta_v$, and the direction of the gradient of $\phi$ as $\theta_\phi$. $\theta_\phi$ is normal to the interface and points towards increasing $\phi$. We quantify anchoring by measuring $\langle\cos2(\theta_i-\theta_\phi)\rangle$, the angle between the director field and concentration gradient, which ranges from $+1$ for homeotropic anchoring to $-$1 for planar anchoring. At interfaces, the contractile fluid tends to align normal to the interface  (Fig.~\ref{fig:Mechanism}(d)) while the extensile fluid prefers to align tangentially (Fig.~\ref{fig:Mechanism}(e)).  We also check the orientation of the relative flow $\delta\mathbf u$ with respect to the normal to the interface by measuring $\langle\cos(\theta_v - \theta_\phi)\rangle$, which is +1 for relative flows pointing towards higher $\phi$, and -1 for relative flows towards lower $\phi$. 
Figure~\ref{fig:Mechanism}(f) shows that the net flow between the fluid components tends to orient normal to the interface, in the direction of increasing concentration gradient. Finally Fig.~\ref{fig:Mechanism}(g) confirms that the magnitude of the relative flow $\langle|\delta \mathbf u|\rangle$ is stronger at interfaces. (See SM Fig.~A1 for a similar figure confirming the same mechanism for imposed alignment between the nematogens, $L_{12} \neq 0$.)

\subsection{No imposed alignment between directors}
\label{sec:PS_FlowAlignment}

In this subsection, we look at phase separation when there is no imposed elastic alignment between the nematic director fields of the two components ($L_{12}=0$). Viscous drag between the fluids aligns (anti-aligns) the directors if the two species have the same (opposite) sign of activity. In all cases, we observe chaotic turbulent microphase separation, with phases forming and dissociating rapidly, similar to Ref. \cite{Bhattacharyya2023}. A snapshot of the concentration field is shown in Fig. \ref{fig:BasicPhaseSeparation}(a) (see also Movie 1 in the SM). 

We characterize the magnitude of phase separation by calculating $\Delta$, the standard deviation of the concentration field, which quantifies the variation from the uniformly mixed state. A contour plot showing how $\Delta$ depends on the activities $\zeta_1$ and $\zeta_2$ is shown in Fig.~\ref{fig:BasicPhaseSeparation}(b). 
The magnitude of phase separation depends on the difference in magnitude of the activities, but not on their signs. 

This is because, although changing the sign of activity changes the relative alignment of the directors,  the flow fields in both fluids remain in the same direction.  Thus the highest phase separation is observed when an active species is mixed with a passive one, corresponding to the maximum relative flows between the two fluids.

\begin{figure*}[htp]
    \centering    
    \includegraphics[width=\textwidth]{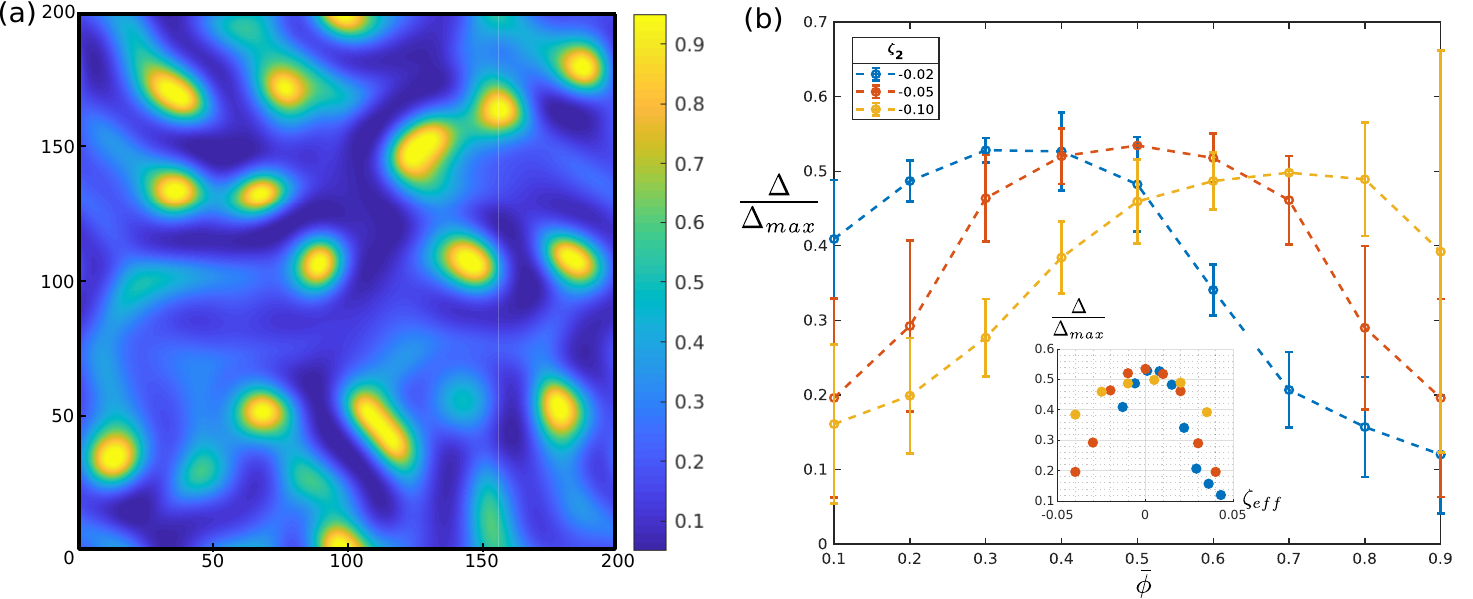}
    \caption{\textbf{Changing Concentration Fraction:} (a) Concentration field for a 30-70 mixture which phase separates into circular droplets ($\zeta_1 = 0.02, \zeta_2 = -0.005,$ and $ \bar\phi=0.3$).   The colourbar shows local concentration $\phi$. (b)  Normalised magnitude of phase separation, $\Delta/\Delta_{max}$,  as a function of concentration fraction, $\bar\phi$, for three different activities $\zeta_2$, while  holding $\zeta_1 = 0.05$ fixed. (Inset) $\Delta/\Delta_{max}$ plotted as a function of effective activity $\zeta_{eff}=\zeta_1 \bar\phi + \zeta_2 (1-\bar\phi)$ is highest at low effective activity.
}
    \label{fig:Conc}
\end{figure*}

\subsection{Imposing director alignment}
\label{sec:PS_ImposedAlignment}

We next impose $L_{12}\neq 0$, so that the director fields of the two nematic components are strongly aligned. A contour plot showing the magnitude of phase separation on varying the activities $\zeta_1$ and $\zeta_2$ is shown in Fig.~\ref{fig:BasicPhaseSeparation}(d).  

If both species have the same sign of activity, the imposed alignment merely reinforces the flow-induced alignment discussed in Sec.~\ref{sec:FlowDrivenAlignment}. The resultant phases look very similar to Fig.~\ref{fig:BasicPhaseSeparation}(a), but the phase ordering is slightly lower because the active flows are better aligned. 

However, the magnitude of the phase separation increases very significantly if the fluid components have opposite signs of activity. The relative flow between the two components is much higher in the extensile-contractile case because each fluid generates active forces in opposite directions at a concentration gradient. The strongest phase separation, corresponding to the largest spread of concentration $\max(\phi) - \min(\phi) \approx 1$, is observed when a highly extensile fluid is mixed with a highly contractile one. 
The fluids form elongated droplet networks with large differences in concentration between the different regions, as shown in Fig.~\ref{fig:BasicPhaseSeparation}(c) (see also Movie 2 in the SM). 

 To directly compare the flow-alignment and imposed alignment cases the variation of the magnitudes of phase separation, $\Delta$, and the relative flow between the fluid components, $\delta \mathbf u$, with $\zeta_1$ for a fixed $\zeta_2$ are plotted in Fig.~\ref{fig:BasicPhaseSeparation}(e) and (f).

\section{Varying other parameters in the system}

\subsection{Concentration Fractions}
\label{sec:Pars_Conc}

We consider the effect of changing $\bar\phi$, the average  of $\phi$ (concentration fraction of species 1), to study how the phase separation is affected if the system is not a 50-50 mixture of each species. 
On reducing $\bar\phi$, the elongated droplet network of Fig.~\ref{fig:BasicPhaseSeparation}(a) is replaced by small droplets of fluid 1 in a background of fluid 2  as shown in Fig.~\ref{fig:Conc}(a) (and Movie 3 in the SM).  At lower activities or higher surface tensions, the droplets are rounder and the flows are less turbulent. 

For a given pair of activities $\zeta_1$ and $\zeta_2$, different concentration fractions $\bar\phi$ give different degrees of phase separation $\Delta$. The maximum possible $\Delta$ for a perfectly phase-separated system with infinitely sharp interfaces is $\Delta_{max} = \sqrt{\bar\phi(1-\bar\phi)}$. Since this is dependent on $\bar\phi$, we quantify the magnitude of phase separation at different $\bar\phi$ using the normalized $\Delta/\Delta_{max}$. 

A plot of $\Delta/\Delta_{max}$ for three different values of $\zeta_2$ is shown in Fig.~\ref{fig:Conc}(b) ($\zeta_1=0.05$). We note that stronger phase separation is achieved when the effective activity $\zeta_{eff} = \zeta_1 \bar\phi + \zeta_2 (1-\bar\phi)$ is small (See inset, Fig. \ref{fig:Conc}(b)) as this suppresses the active turbulence which leads to droplets breaking up.
 This implies that  the highest phase separation is observed when the more active component has a smaller concentration fraction. 

\subsection{Friction}
\label{sec:Pars_Fric}

\begin{figure*}[htp]
    \centering    
    \includegraphics[width=\textwidth]{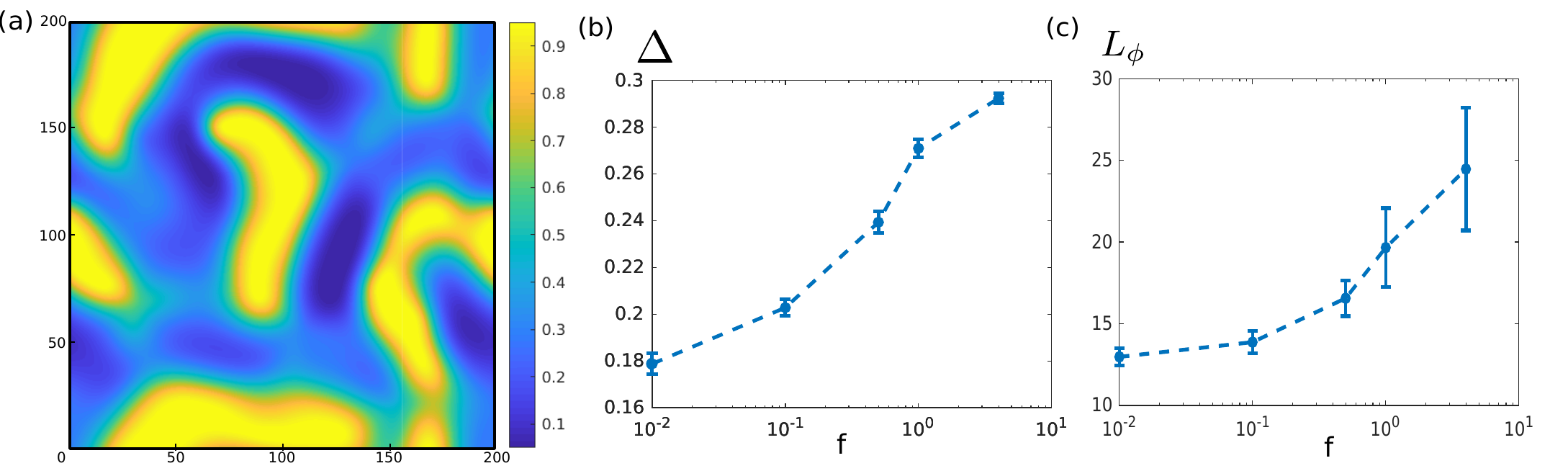}
    \caption{\textbf{Changing Substrate Friction:} (a) 
 Concentration field for high substrate friction ($\zeta_1 = 0.05, \zeta_2 = -0.05, f=4$), showing large, well-separated domains. The colourbar shows local concentration $\phi$.  (b) Magnitude of phase ordering,  $\Delta$, increases with higher substrate friction. (c) Size of phase-separated regions, $L_\phi$, increases significantly with higher substrate friction. }
    \label{fig:Friction}
\end{figure*}

Finally, we look at the effect of adding substrate friction $f$.
On increasing friction, the active flows are weaker and less turbulent. The elongated droplet networks tend to aggregate  to form more strongly phase-separated regions that are larger in size.  A snapshot of the concentration profile  is shown in Fig.~\ref{fig:Friction}(a) (see also Movie 4 in the SM). Fig.~\ref{fig:Friction}(b) shows the increase in the magnitude of phase separation upon increasing the substrate friction. The typical lengthscales of the phase-separated domains, $L_\phi$, also increase with higher substrate friction (Fig.~\ref{fig:Friction}(c)).

At first glance, it might appear surprising that high substrate friction facilitates phase separation driven by active flows. However, although the overall dynamics are slower, the active turbulent flows are smaller in magnitude, and dissociation by active instabilities is very rare. Moreover, the force balance condition changes and the velocity is now directly proportional to the applied force $\mathbf v^i \approx \mathbf F^i / f$, instead of $\mathbf \nabla^2 v^i \approx \mathbf F^i / \eta$. As a result, although the velocity field is weaker, it is better aligned at the interface, as shown in Fig. A2 in the SM. 

\section{Discussion}

To summarise, we have extended a two-fluid model to study mixtures where both species are active nematics. We argued that even in the absence of any externally imposed elastic ordering between components, active nematics coupled by viscous drag tend to align or anti-align  depending on the relative signs of the activities and the value of the flow-tumbling parameter. 

 We observed turbulent microphase separation in mixtures of two active nematic fluids, each with a different activity. Imposing an elastic director alignment between the two active species plays a major role in determining the magnitude of the phase separation. In the absence of imposed alignment, there is weak segregation which is most pronounced for a mixture of a highly active and a passive component. However, in the presence of imposed alignment, coexisting domains comprised almost entirely of one component can be achieved with an extensile--contractile fluid mixture. 

 The degree and morphology of phase separation are also affected by other parameters in the system. As expected, varying the concentration fraction can change the morphology from elongated droplet networks to isolated circular drops. Perhaps more surprisingly, increasing substrate friction leads to stronger phase separation and a considerable increase in the size of the phase-separated domains. 

Our results add to a growing body of work describing the possibility of phase separation driven by activity,  here focusing on active flows in fluid mixtures. Cell ordering and sorting in embryogenesis, and compartmentalisation into different cell types, are important biological processes where activity may play a role \cite{Balasubramaniam2022, Skamrahl2023, Graham2024, Krajnc2020, Heine2021}. Our work may also be relevant in understanding organization and movement in multi-species bacterial colonies \cite{Hallatschek2023, DellArciprete2018}, or the intracellular liquid-liquid phase separation \cite{AgudoCanalejo2019, Hyman2014} which results in membrane-less organelles.

With biological examples in mind, in future work, it will be interesting to consider nematic mixtures above the nematic-isotropic transition temperature, compare phase ordering in the vertex model or multiphase-field description of cells, and study the interplay between confinement, wetting and phase ordering in active materials.

\begin{acknowledgements}
We thank A. Mietke, J. Rozman, J. N. Graham, and I. Hadjifrangiskou for valuable discussions, and J. Rozman for helpful comments on the manuscript. SB acknowledges support from the Rhodes Trust and the Crewe Graduate Scholarship. JMY acknowledges support from the UK EPSRC  (Award EP/W023849/1).
\end{acknowledgements}

\bibliography{refs}

\end{document}